\documentclass[twocolumn,preprintnumbers,superscriptaddress,nofootinbib,aps,prd,floatfix]{revtex4}

\usepackage{amsmath,amssymb}
\usepackage{graphicx}
\usepackage{subfigure,slashed}
\usepackage{color}
\usepackage{multirow}
\usepackage{hyperref}

\usepackage{bbold}

\usepackage{array}
\newcolumntype{L}[1]{>{\raggedright\let\newline\\\arraybackslash\hspace{0pt}}m{#1}}
\newcolumntype{C}[1]{>{\centering\let\newline\\\arraybackslash\hspace{0pt}}m{#1}}
\newcolumntype{R}[1]{>{\raggedleft\let\newline\\\arraybackslash\hspace{0pt}}m{#1}}

\newcommand{\be}{\begin{eqnarray*}}
\newcommand{\ee}{\end{eqnarray*}}

\newcommand{\bee}{\begin{eqnarray}}
\newcommand{\eee}{\end{eqnarray}}
\newcommand{\beeq}{\begin{equation}}
\newcommand{\eeeq}{\end{equation}}

\DeclareMathOperator{\arcsinh}{arcsinh}
\newcommand{\dd}{\, {\rm d}}
\newcommand{\mmm}{{_{\rm m}}}
\newcommand{\tg}{\tilde{g}}
\newcommand{\nm}{{\mu\nu}}

\newcommand{\eff}{_{\rm eff}}
\newcommand{\ms}{M}
\newcommand{\rst}{\rho_\star}

\begin{document}

\title{Detecting Dark Domain Walls}
\begin{abstract}
Light scalar fields, with double well potentials and direct matter couplings, undergo density driven phase transitions, leading to the formation of domain walls. Such theories could explain dark energy, dark matter or source the nanoHz gravitational-wave background. We describe an experiment that could be used to detect such domain walls in a laboratory experiment, solving for the scalar field profile, and showing how the domain wall affects the motion of a test particle. We find that, in currently unconstrained regions of parameter space, the domain walls leave detectable signatures.
\end{abstract}

%
%

\author{Kate Clements} 
\email{Kate.Clements@nottingham.ac.uk}
\affiliation{School of Physics and Astronomy, University of Nottingham, Nottingham, NG7 2RD, United Kingdom\\[0.2cm]}

\author{Benjamin Elder}
\email{bcelder@hawaii.edu}
\affiliation{Department of Physics and Astronomy, University of Hawai'i, 2505 Correa Road, Honolulu, HI 96822, USA\\[0.2cm]}

\author{Lucia Hackermueller} 
\email{Lucia.Hackermuller@nottingham.ac.uk}
\affiliation{School of Physics and Astronomy, University of Nottingham, Nottingham, NG7 2RD, United Kingdom\\[0.2cm]}

\author{Mark Fromhold} 
\email{Mark.Fromhold@nottingham.ac.uk}
\affiliation{School of Physics and Astronomy, University of Nottingham, Nottingham, NG7 2RD, United Kingdom\\[0.2cm]}

\author{Clare Burrage} 
\email{Clare.Burrage@nottingham.ac.uk}
\affiliation{School of Physics and Astronomy, University of Nottingham, Nottingham, NG7 2RD, United Kingdom\\[0.2cm]}

\preprint{}
\maketitle

\section{Introduction}
\label{sec:intro}
 Light scalar fields are commonly invoked to solve cosmological mysteries, including the nature of dark matter \cite{Hu:2000ke,Ferreira:2020fam},   the source of the current accelerating expansion of the universe \cite{Copeland:2006wr,Joyce:2014kja}, or whether gravity is modified on the largest cosmological scales \cite{Clifton:2011jh}.  Such scalar fields can arise in string theory and other attempts to complete the Standard Model at high energies \cite{Damour:2002nv,Gasperini:2001pc}, and can  be introduced into low-energy effective field theory descriptions of physics in natural ways \cite{Binoth:1996au,Patt:2006fw,Schabinger:2005ei,Englert:2011yb,Bauer:2020nld,Beacham:2019nyx}. 

Adding a scalar field is relatively simple at the level of the theoretical Lagrangian, and a wide range of interesting phenomenology can result depending on the choice of potential and couplings. Allowing scalar potentials beyond just a simple mass term can weaken  experimental constraints due to screening \cite{Joyce:2014kja,Slosar:2019flp,Brax:2021wcv}, but also yield  novel detectable signatures.  In this work we focus on scalar fields that couple to matter and have a symmetry breaking potential, which allows for the formation of domain wall topological defects when the local energy density is lowered below a critical threshold \cite{Hinterbichler:2010es,Hinterbichler:2011ca} (for related work see Refs.  \cite{Mota:2006fz,Dehnen:1992rr, Gessner:1992flm, Damour:1994zq, Pietroni:2005pv, Olive:2007aj}). 

Domain walls are planar topological defects which store energy and form after the scalar field goes through a phase transition.  There has been much study of the observable consequences of scalar fields undergoing temperature driven phase transitions in the early universe \cite{Vilenkin:1984ib,Lazanu:2015fua}.  Here, we focus on an alternative possibility that, due to direct couplings to matter, the phase transitions are  driven by changes in energy density, and that this could lead to detectable effects in laboratory experiments \cite{Brax:2021wcv}. The coupling to matter changes the scaling of a cosmological network of domain walls \cite{Llinares:2014zxa,Pearson:2014vqa}, and they may be unstable on cosmological timescales \cite{Christiansen:2020lyy}. We call these `dark' domain walls, because the coupling of the scalar field to matter will, in many scenarios, be small and so difficult to see unless experiments and observables are carefully tailored.

A direct coupling of the scalar field to matter has many implications. For this work we will be particularly concerned with three: 1.~The scalar field mediates a fifth force, whose strength  is proportional to the background value of the scalar. 2.~Symmetry breaking is controlled by the local matter density. 3.~Once formed the domain walls can be `pinned' to matter structures. 
A consequence of the first  property is that matter particles passing through a domain wall can be trapped, or deflected \cite{Llinares:2014zxa,Naik:2022lcn}.  
The possibility of detecting topological defects through these effects has been considered in \cite{Llinares:2018mzl} in the context of experiments with Ultra-Cold Neutrons. On a different scale, a dark domain wall of the type discussed in this work could also provide an explanation for the observed planes of satellite galaxies around the Milky Way and Andromeda \cite{Naik:2022lcn}. A network of cosmological domain walls has also been proposed as a source for the recently observed nanoHz stochastic background of gravitational waves \cite{Nakayama:2016gxi,Saikawa:2017hiv,Ferreira:2022zzo}.

In this work we  explore the conditions needed for  domain walls to form inside a laboratory vacuum chamber, and the implications for the matter structures needed to pin the domain walls in place, enabling possible detection. We show that the Kibble-Zurek mechanism alone is not sufficient to facilitate the formation of domain walls as the gas density changes inside a vacuum chamber, but that their formation can be encouraged with a suitably designed experiment.  We demonstrate that there exists a region of currently unconstrained parameter space, within which  such domain walls could cause measurable deflections of clouds of cold atoms. We work with a $(-,+,+,+)$ metric and use natural units unless otherwise stated.

\section{The model}
We consider a theory where the Standard Model is supplemented by a scalar field with a symmetry breaking potential, and  the scalar field interacts directly  with the standard model fields.  This can equivalently \cite{Burrage:2018dvt} be thought of as  a  scalar-tensor theory, or a scalar coupled to the Standard Model through the Higgs portal.  

As a scalar tensor theory the theory has the form
\begin{align}
\label{eq:STtheoriesgeneral}
S=&\int\dd^4 x\sqrt{-g}\left[\frac{R}{16\pi G}-\frac{1}{2}\nabla_\mu\phi\nabla^\mu\phi-V(\phi)\right]\nonumber\\
&+S\mmm[\tg_{\mu\nu},\psi_i]\;,
\end{align}
where the scalar field is conformally coupled to matter fields, $\psi_i$, through the Jordan frame metric $\tg_\nm=A^2(\phi)g_\nm$ and the matter action $S\mmm$. 
 The bare potential\footnote{One might worry about radiative corrections spoiling the shape of this potential, however in Ref. \cite{Burrage:2016xzz} a model was constructed, based on the multi-field model of  \cite{Garbrecht:2015yza}, where the one loop potential can undergo a symmetry breaking transition whilst the higher order loop corrections remain small. } and coupling function are
\begin{equation}
V(\phi)=-\frac{1}{2}\mu^2\phi^2+\frac{\lambda}{4}\phi^4;\quad A(\phi)=1+\frac{\phi^2}{2M^2}\;,
\end{equation}
where $\mu$ and $M$ are constant mass scales, and $\lambda$ is a positive dimensionless constant.
A test particle experiences a  non-relativistic fifth force mediated by the scalar field of 
\begin{equation}
\label{eq:F5gen}
\mathbf{F}_5=-\mathbf{\nabla} A(\phi) \approx -\frac{\phi \mathbf{\nabla} \phi}{M^2}.
\end{equation}

The scalar equation  of motion is
\begin{equation}
\label{eq:EOMgen}
\Box\phi=\frac{\dd V(\phi)}{\dd\phi}-\frac{\phi T}{M^2}\;,
\end{equation}
where $T=g_{\mu\nu}T^{\mu\nu}$ and $T^\nm=2/\sqrt{-g}\delta S\mmm/\delta g_\nm$ is the Einstein frame energy-momentum tensor.  For non-relativistic matter, one has $T=-\rho$, where $\rho$ is the energy-density. We can  
 define a density-dependent effective potential for the scalar field:
\begin{equation}
\label{eq:veff_symmetron}
V\eff(\phi)=-\frac{1}{2}\mu^2\left(1-\frac{\rho}{\mu^2\ms^2}\right)\phi^2+\frac{\lambda}{4}\phi^4\;.
\end{equation}
The coefficient of the  term quadratic in $\phi$ can be either positive or negative depending on whether the density is above or below the  critical density, defined as 
\begin{equation}
\rst \equiv \mu^2 M^2.
\label{eq:rhostar}
\end{equation}

The model described here contains three free parameters; $\mu$, $M$ and  $\lambda$ and different choices explain different observed phenomena.  For example: When $\rst$ coincides with the present day cosmological density it may help explain the late time dark energy domination of our universe \cite{Hinterbichler:2010es,Hinterbichler:2011ca}.  When $\mu M_{\rm pl}= \sqrt{\lambda} M^2$ the fifth force in vacuum is of gravitational strength, suggesting a connection to theories of modified gravity. When $\mu \approx 10^{-27} \mbox{ eV}$ and $ \sqrt{\lambda}M \approx 10^{-24} \mbox{ eV}$ the model can  explain the observed planes of Milky Way satellite galaxies \cite{Naik:2022lcn}. When $\mu/\lambda^{1/3} \approx \mbox{MeV}$ domain walls can make up a fraction of the dark matter density in the universe today  \cite{Stadnik:2020bfk}. When  $\mu/\lambda^{1/3} \approx 10^5 \mbox{ GeV}$, the  domain walls could source  the observed stochastic gravitational wave background \cite{Ferreira:2022zzo}.

Current constraints on the model parameters (for a fixed value of the dimensionless constant $\lambda$) are shown by the dark opaque shaded region in Figure \ref{fig:region}. The parameter space shown in this figure corresponds to domain walls with a critical density that means they would form between the epoch of Big Bang Nucleosynthesis and the time at which the universe becomes matter dominated. 

 \section{Infinite Domain walls}
\label{sec:infinite}
At low densities the scalar potential in equation (\ref{eq:veff_symmetron}) has two degenerate minima.  This allows for the formation of domain walls; topological defects whose field profile smoothly interpolates between these two minima.
Infinite, straight, static domain walls can be studied analytically and have the profile
\begin{equation}
\phi(y) =\phi_0\tanh (y/d)\;,
\label{eq:profile}
\end{equation}
where $d$, the width of the domain wall, is given by $ d=\frac{\sqrt{2}}{\mu}$ and 
\begin{equation}
    \phi_0^2 =\frac{\mu^2}{\lambda}\left(1-\frac{\rho_0}{\rst}\right)\;.
\end{equation}

The possibility of detecting scalar domain walls through their impact on the trajectories of matter particles has been considered in many contexts from cosmology \cite{Vilenkin:2000jqa} through solar system dynamics \cite{Dai:2021boq} to laboratory experiments \cite{Llinares:2018mzl}.  
In \cite{Llinares:2018mzl} the trajectory of a test particle moving perpendicularly through an infinite,  straight domain wall was computed. The particle dynamics can be expressed in terms of the conserved Hamiltonian of the test particle, assumed to have unit mass, 
\begin{equation}
    H_y = \frac{\dot{y}^2}{2}+ \frac{\phi^2(y)-\phi_0^2}{2M^2}\;.
\end{equation}
 The  behaviour of the particle is governed by the sign of the Hamiltonian; for an infinite straight domain wall, this means that the critical initial conditions  separating the two regimes of behaviour satisfy
\begin{equation}
\cosh\left(\frac{y_{\rm{crit}}(t=0)}{d}\right) = \frac{\phi_0}{\dot{y}_{\rm{crit}}(t=0) M}
\label{eq:xiest}
\end{equation}
where $y_{\rm{crit}}(t=0)$ is the critical initial position and $\dot{y}_{\rm{crit}}(t=0)$ is the critical initial velocity. 

We see two types of behavior for the test particle:
\begin{itemize}
    \item If $a^2 = 1+\phi_0^2/(2H_y M^2)>0$ the particle passes through the domain wall and 
\begin{align}
\sinh\left(\frac{y(t)}{d}\right) = a \sinh& \left(\frac{(2H_y)^{1/2}(t-t_0)}{ d}\right.\nonumber\\
&+\left.\arcsinh \left(\frac{\sinh(y_0/d)}{a}\right)\right)\;.
\end{align}

\item If $\alpha^2 = -[1+\phi_0^2/(2H_y M^2)]>0$ the particle gets trapped within the domain wall and
\begin{align}
\sinh\left(\frac{y(t)}{d}\right) = \alpha \sin & \left(\frac{(-2H_y)^{1/2}(t-t_0)}{ d}\right.\nonumber\\
&+\left.\arcsin \left(\frac{\sinh(y_0/d)}{\alpha}\right)\right)\;,
\end{align}
\end{itemize}
where $y_0=y(t_0)$ and $\dot{y}_0 =\dot{y}(t_0)$. 

The perturbation, $\Delta y$, caused by a domain wall  to the motion of a particle that would otherwise move  at constant speed is shown in Figure \ref{fig:region} for  $\lambda=10^{-10}$. Smaller values of $\lambda$ give rise to stronger fifth forces and larger displacements, and larger $\lambda$ gives smaller displacements. For reference, displacements of $10\;\mu\mbox{m}$ or more can be detected with modern high precision cameras. 

\section{Detecting Dark Walls}
In order to design a laboratory experiment that is able to detect dark domain walls, we must first ensure that domain walls can form. We consider an idealised experiment performed inside a vacuum chamber whose walls have a fixed density, and where the gas pressure (and thus density) inside the chamber can be varied.   A necessary condition for domain walls to form is that the density of gas can be decreased with time through the critical density $\rst$ (defined in Eq.~(\ref{eq:rhostar})). We also require the density of the walls of the vacuum chamber to be  above $\rst$, so that the effective mass of the field inside the walls is  large, and  perturbations of the scalar field sourced outside the  chamber are  exponentially damped in the walls, and can be neglected. We therefore   require
\begin{align}
5.6 \times 10^{-5}\left(\frac{\rho_{\rm gas}}{10^{-18} \mbox{ g/cm}^{3}} \right)< & \left(\frac{ M}{\mbox{GeV}}\right)^2 \left(\frac{\mbox{m}}{d}\right)^2 \label{eq:rhobound}\\
&< 5.6 \times 10^{13}\left(\frac{\rho_{\rm wall}}{ \mbox{g/cm}^{3}}\right)\;,\nonumber 
\end{align}
where $\rho_{\rm gas}$ is the minimum value of the gas density inside the chamber. We also require that the width of the domain wall be smaller than the characteristic internal dimension of the vacuum chamber $L$; $   d< L$.

\begin{figure}[t!]\centering
\includegraphics[width=0.5\textwidth]{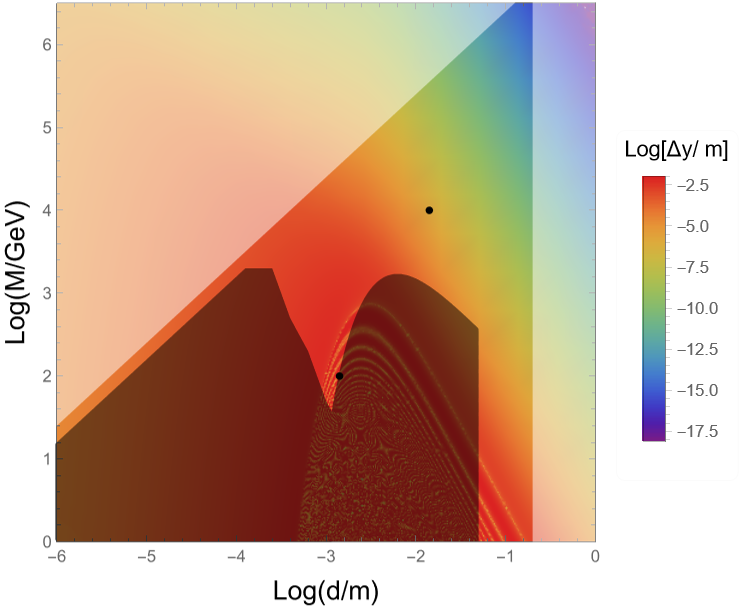}
\caption{The parameter space for the scalar field theory, with $\lambda =10^{-10}$. The dark opaque
shaded region is excluded by existing constraints from neutron bouncing experiments, cold neutron interferometry \cite{Jenke:2020obe}, and atom interferometry \cite{Sabulsky:2018jma}. The light opaque shaded regions show where domain walls would not form within our idealised vacuum chamber.  The color scale shows the expected deviation from linear motion of a test particle, $\Delta y = y(t) - y_0+ \dot{y}_0t$, with initial velocity $\dot{y}_0=10^{-3}\mbox{ m/s}$, and starting position $y_0=5\times10^{-3}\mbox{ m}$ moving perpendicularly to an infinite straight domain wall located at $y=0$, after $10\mbox{ s}$. Towards the bottom right of the plot we see the oscillations (lighter coloured curved domains) in position caused by the test particle being trapped by the domain wall. Two black dots show the example points in parameter space discussed in Section \ref{sec:obs}.}\label{fig:region}
\end{figure}

In this work we consider an idealised experiment, where a spherical vacuum chamber has an internal radius $L= 10\mbox{ cm}$. The density of the stainless steel walls of the chamber is $\rho_{\rm wall}=8\mbox{ g/cm}^{3}$ and the minimum vacuum pressure is $10^{-11}\mbox{ mbar}$ (corresponding to a vacuum gas density at room temperature  $\rho_{\rm gas}\approx 10^{-18}\mbox{ g/cm}^{3}$). The upper bounds on $d$ and $M/d$ required for domain walls to form in such an environment are shown respectively by the light opaque shaded rectangular and  triangular regions in Figure \ref{fig:region}.

\subsection{Formation of domain walls}

If the density of the gas in the vacuum chamber is lowered through $\rst$ a symmetry breaking transition occurs.  This is a necessary condition for domain walls to form but not a sufficient one. If the gas density inside the vacuum chamber is lowered uniformly then domain walls can form through the Kibble-Zurek mechanism \cite{Kibble:1976sj,Zurek:1985qw,Zurek:1996sj}. Assuming that the gas density varies  linearly with time through the phase transition such that $|\rho(t)-\rst|/\rst = t/T$, for some characteristic timescale $T$, then we expect that domain walls will start to form at time $\hat{t} \approx (T/\mu^2)^{1/3}$ (when evolution of the scalar field is no longer adiabatic) and with correlation length $L_c= d \left(T/2d\right)^{1/3}$.
 If  the correlation length  is smaller than the characteristic size of the vacuum chamber, $L_c<L$, then domain walls may form inside the chamber.   This requires (reintroducing physical units)
\begin{equation}
\left(\frac{d}{\mbox{m}}\right)^{2} \lesssim 10^{-12}\left(\frac{L}{\mbox{m}}\right)^3\frac{\mbox{sec}}{T}\;.
\end{equation}

The  timescale for lowering the gas density of a vacuum  chamber varies. If, for example,  $T = 10^{-1} \mbox{s}$ then domain walls with $d \lesssim 2.7\times10^{-5} \mbox{ m}$ will have a correlation length smaller than the size of the vacuum chamber, $0.1 \mbox{ m}$, meaning that we can expect at least one domain wall to form inside the volume. From Figure \ref{fig:region} we see that there is little  parameter space available for such thin domain walls to form in our idealised experiment that has not already been excluded.  Thicker domain walls will be unlikely to form inside the spherical chamber through the Kibble-Zurek mechanism alone.

The Kibble mechanism describes the formation of domain walls in infinite space, but we are considering an experiment inside a finite sized vacuum chamber. Structures inside the vacuum chamber can both encourage domain walls to form, and  influence where they form. If spikes protrude from the walls of the chamber, such that  the space between the tips of the  spikes is smaller than, or similar to, the Compton wavelength $d$ of the scalar field, then the field will not have space to change its value from zero, even as the density of the chamber is lowered. It has been shown previously \cite{Llinares:2014zxa,Pearson:2014vqa,Llinares:2018mzl} that matter structures with even larger separations can be used to stabilise domain walls and pin them in place. As a result, the presence of spikes protruding from the walls of a vacuum chamber make it extremely likely that a non-trivial scalar field profile is present.\footnote{The field may not choose different vacua on either side of the spikes. Different points inside the vacuum chamber choose a vacuum value for the scalar field at random, so the experiment would need to be repeated multiple times  to increase the probability that a domain wall forms in at least one iteration of the experiment.} An example of such a configuration is shown in Figure \ref{fig:thin_trapped_field}, where we assume rotational symmetry around the $y$ axis. Simulation of the time dependent behavior of the field during this process is left for future work.

To further encourage a domain wall to form,  a shutter may be placed in the  narrow waist between the spikes protruding from the vacuum chamber walls.  While the density of the vacuum gas is lowered, the  shutter is closed,  allowing the field to evolve separately on either side.  In each region the field  has an equal chance of rolling to either of the two vacuum expectation values (vevs) as the density of the vacuum gas decreases.  Consequently, there is a 50\% chance that the field rolls to different vevs in the two chambers.  Once the vacuum chamber is fully pumped out, the shutter is opened quickly.  In the event that the field has rolled to a different minimum in each of the lobes, the field in the region of the waist will interpolate between those two vevs, and a  domain wall will be present.

\subsection{Observable signatures of domain walls}
\label{sec:obs}
We consider the effects of a domain wall on the motion of a test particle falling freely through the wall.   We use a finite element code adapted from SELCIE \cite{Briddon:2021etm}, which can solve static, non-linear scalar field equations around arbitrary-shaped matter configurations, to determine the scalar field profile.  We solve for the motion of a test particle in this scalar field background using a leapfrog algorithm. These codes are described  in the supplementary material.   

We consider two example cases, a thin and a thick domain wall, which correspond to two currently-unconstrained points in parameter space indicated in Figure \ref{fig:region}. For both scenarios we keep the geometry of the chamber  fixed.  In the case of the thin domain wall, this allows us to test the results of our code against the analytic predictions of Section \ref{sec:infinite}, and in the case of the thick domain wall to see the effects of the vacuum chamber geometry on the domain wall profile and its resulting influence on the motion of particles.

\subsubsection{Thin domain walls}

\begin{figure}[t!]\centering
\includegraphics[width=0.5\textwidth]{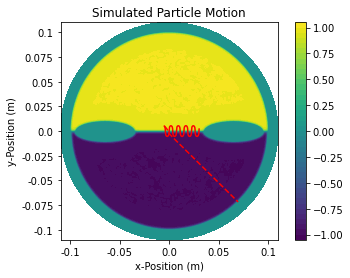}
\caption{The motion of two particles with initial position $\mathbf{x}_0=(-0.005,0.005)\mbox{ m}$ and initial velocities $\mathbf{\dot{x}}_0=(0.001, -0.001) \mbox{ m/s}$ (solid line) and $(0.01, -0.01)\mbox{ m/s}$ (dashed line) in a vacuum chamber. The solid line shows the motion of a trapped particle, and the dashed line shows a particle being deflected. Scalar parameters are $\mu = 2.0 \times 10^{-13}\mbox{ GeV}$, $M=100\mbox{ GeV}$ and $\lambda=10^{-10}$ (lower black dot in Figure \ref{fig:region}). 
The color bar indicates the value of the scalar field, with $\phi_0$ normalised to 1. The simulation runs for $t = 35\text{s}$.}\label{fig:thin_trapped_field}
\end{figure}

A thin domain wall forms  in a theory with $\mu = 2.0 \times 10^{-13}\mbox{ GeV}$, $M=100\mbox{ GeV}$ and $\lambda=10^{-10}$. As the width of the domain wall is much smaller than the internal dimensions of the vacuum chamber  the effects of the domain wall are well modeled by the analytic approximation of an infinite straight domain wall. 

For a particle starting at a point $y_0=5\times 10^{-3} \mbox{ m}$ above the centre of the domain wall, we see from the analytic estimate in equation (\ref{eq:xiest}) that  particles with initial velocities below $\dot{y}_0=3 \times 10^{-2} ~\mbox{m/s}$ will be trapped by the domain wall, and those with higher initial velocities will pass through the wall. This agrees with the results of our numerical simulations, shown in Figure \ref{fig:thin_trapped_field}. A particle with initial velocity $\dot{y}_0=1 \times 10^{-2} ~ \mbox{m/s}$ is  trapped by the wall whereas a particle with initial velocity $\dot{y}_0=1 \times 10^{-1} ~ \mbox{m/s}$ passes through. 

This example allows us to check the validity of the analytic approximation of equation (\ref{eq:xiest}) by comparing the horizontal and vertical motions for a particle that is trapped by the domain wall. If the initial horizontal  position and velocity are $x_0=-5\times 10^{-3}~\mbox{ m}$  and $\dot{x}_0=1 \times 10^{-3} ~\mbox{m/s}$ and there is no acceleration of the particle in the $x$ direction,  the particle will have position  $x=+5 \times 10^{-3}~ \mbox{m/s}$ after $10\mbox{ s}$.  Numerical evolution of the particle motion reproduces this motion with an accuracy of $0.04 \%$. In contrast, the domain wall  causes an acceleration in the vertical direction. After $10\mbox{ s}$ the analytic calculation predicts that the position of the particle is $y=-5.78 \times 10^{-3} \mbox{ m}$, while our numerical integration finds a position of $y=-5.30 \times 10^{-3} \mbox{ m}$, a difference of $8\%$.  So  even when the domain wall is thin compared to the size of the chamber, the experimental environment needs to be simulated in order to produce an accurate prediction of the motion of test particles. 

\subsubsection{Thick domain walls}

\begin{figure}[t!]\centering
\includegraphics[width=0.5\textwidth]{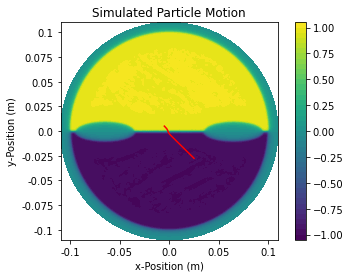}
\caption{The motion of a particle with initial position $\mathbf{x}_0=(-0.005,0.005) \mbox{ m}$ and initial velocity $\mathbf{\dot{x}}_0=(10^{-5}, -10^{-5}) \mbox{ m/s}$ in a vacuum chamber. Scalar parameters are $\mu = 2.0 \times 10^{-14}\mbox{ GeV}$, $M=10^4\mbox{ GeV}$ and $\lambda=10^{-10}$ (upper black dot in Figure \ref{fig:region}). 
The color bar indicates the value of the scalar field, with $\phi_0$ normalised to 1. The simulation runs for $t = 3000\text{s}$.}\label{fig:thick_pass_field}
\end{figure}

A thick domain wall forms   in a theory with $\mu = 2.0 \times 10^{-14}\mbox{ GeV}$, $M=10^4\mbox{ GeV}$ and $\lambda=10^{-10}$. We do not expect the effects of such  domain walls to be well modeled by the analytic approximation of an infinite straight domain wall. 
In this case, for the values of the initial velocity that we have simulated (down to $10^{-5}\mbox{ m/s}$) we find that the particle always passes through the wall. An example  trajectory can be seen in Figure \ref{fig:thick_pass_field}.

Considering the motion of a particle starting at a point $y_0=5\times 10^{-3} \mbox{ m}$   we find that, at the accuracy of our simulation, the effects of the domain wall are only visible for initial velocities $\dot{y}_0=10^{-5} \mbox{ m/s}$ or lower.  For a particle with this initial velocity we find that after a period of $10^3 \mbox{ s}$ ($16.7 \mbox{ mins}$) the particle's position is $y=-7.7\times 10^{-3}\mbox{ m}$, whereas, with no  domain wall the particle's position would be $y=-5 \times 10^{-3} \mbox{ m}$. The analytic prediction, modeling the domain wall as straight and infinite,  predicts the position at this time to be $y= -1.2 \times 10^{-3} \mbox{ m}$, a difference that would be detectable with a high resolution camera.

\section{Conclusion}
Light scalar fields are increasingly of interest as they can form part or all of the dark matter and dark energy needed to complete our cosmological model. While much work currently considers scalar fields with simple quadratic potentials for the light scalar, more complex potentials, such as symmetry breaking potentials, are well motivated, and give rise to different observables.

Domain walls are a signature of symmetry breaking potentials for real scalar fields and, when the scalar field couples to matter, test particles passing through a domain wall will experience additional forces mediated by the scalar.  When initial velocities are low, the particles can become trapped within the potential well of the domain wall, and when initial velocities are larger the trajectory of the particle can be deflected by the domain wall.

The possibility of detecting such domain walls in laboratory experiments has been previously discussed, but the probability of forming a domain wall in such an experiment has not.  In this work we have argued that structures inside the experimental vacuum chamber, produced for example by 3D printing \cite{COOPER2021101898}, can be exploited to pin domain walls in place, otherwise the domain walls are unlikely to form or will be short lived. In the presence of these structures it is necessary to solve for the scalar field profile inside the vacuum chamber numerically.  We have done this inside an idealised chamber, for a selection of model parameters. We expect our results to translate to realistic experiments.  For example, although we have assumed a spherical vacuum chamber with protruding spikes, the key feature is just that there are two vacuum regions, larger than the Compton wavelength of the scalar field,  within which the field can reach its minimum value. 

We find that, across a range of currently unconstrained parameter space, domain walls can give rise to observable deflections in particle motion, including the  trapping  of test particles within the domain wall. This opens up exciting possibilities for relatively simple experiments using  test particles, which could be cold atoms or molecules, or nanobeads, to detect or constrain this beyond-the-standard-model physics. 

\section*{Acknowledgements}
CB is supported by a Research Leadership Award from the Leverhulme Trust and STFC grant ST/T000732/1. 
LH and MF acknowledge support by IUK projects No. 133086 and 10031462, the EPSRC grants EP/R024111/1, EP/T001046/1, EP/Y005139/1 and the JTF grant No. 62420.

\appendix

\section{Simulating the field}
\label{app:sim}
The equations of motion governing the behaviour of our scalar field are non-linear, and to have the capacity to solve them within experimental environments requires a numerical simulation. The code we use to do this is based on a modification of the numerical code SELCIE \cite{Briddon:2021etm} which solves the equations of motion for screened scalar field theories using the finite element approach.

SELCIE uses the open-source finite element software FEniCS. It returns profiles for the chameleon field \cite{Khoury:2003rn}, a non-linear scalar tensor theory with different choices of potential and coupling function to those we consider in this work, which does not give rise to domain walls. We have written a new code that takes the user-defined mesh from SELCIE and finds field profiles for the symmetron field instead.

We consider time-independent solutions, so that the symmetron equation of motion is:
\begin{equation}
    \nabla^2\phi = \left(\frac{\rho}{M^2} - \mu^2\right)\phi + \lambda\phi^3\;.
    \label{symmetron_eom_diff_form}
\end{equation}
In order to eliminate $\lambda$ from the equation of motion, we introduce a new field variable $\hat{\phi}\equiv\phi/\phi_0$ 
and a new  mass scale $\hat{\mu}^2 = \frac{M^2}{\rho_{ 0}}\mu^2$, where $\rho_{0}$ is a reference density, typically taken to be the density of the vacuum gas. 
We also rewrite the Laplacian such that $\hat{\nabla}^2\equiv L^2\nabla^2$ where $L$ is the length scale of the vacuum chamber. Substituting these new definitions into equation (\ref{symmetron_eom_diff_form}), we find
\begin{equation}
    \alpha\hat{\nabla}\hat{\phi} = -\left(\hat{\mu}^2-\hat{\rho}\right)\hat{\phi} + \left(\hat{\mu}^2-1\right)\hat{\phi}^3\;,
    \label{rescaled_symmetron_eom}
\end{equation}
where $\hat{\rho}\equiv\rho/\rho_{0}$ and the dimensionless constant $\alpha$ is given by
\begin{equation}
    \alpha = \frac{M^2}{L^2\rho_0}\;.
\end{equation}
\subsubsection{Matrix version of the symmetron equation of motion}
In order to input the symmetron equation of motion into the SELCIE code, we must write it as a matrix equation that can be solved iteratively. The first step to finding the matrix representation for the symmetron equation of motion is to write equation (\ref{rescaled_symmetron_eom}) in integral form. Following the method in \cite{Briddon:2021etm}, we multiply both sides of equation (\ref{rescaled_symmetron_eom}) by a test function $v_j$,
and integrate over the domain $\Omega$ to find:
\begin{align}
       \alpha\int_\Omega\hat{\nabla}\hat{\phi}v_jdx =& -\int_\Omega\left(\hat{\mu}^2-\hat{\rho}\right)\hat{\phi}v_jdx \nonumber\\
       &+ \int_\Omega\left(\hat{\mu}^2-1\right)\hat{\phi}^3v_jdx\;.
       \label{integral_omega}
\end{align}
The left-hand-side of the above equation can be integrated by parts, and we find that the resulting 
 boundary term vanishes if we choose $v_j$ to vanish on $\partial\Omega$, the boundary of the domain,  for all $j$. This results in the integral form of the rescaled symmetron equation of motion:
    \begin{align}
    \alpha\int_\Omega\hat{\nabla}\hat{\phi}\cdot\hat{\nabla}v_jdx =& \int_\Omega\left(\hat{\mu}^2-\hat{\rho}\right)\hat{\phi}v_jdx \nonumber\\
    &- \left(\hat{\mu}^2-1\right)\int_\Omega\hat{\phi}^3v_jdx\;.
    \label{symmetron_eom_integral}
    \end{align}

We use  the Picard method to solve the equation of motion in the following way. We  expand  the non-linear term in the equation of motion around some field value $\hat{\phi}_k$, which will be updated at every iteration of the solver. The linearised equation of motion is then 
\begin{align}
    \alpha\int_\Omega\hat{\nabla}\hat{\phi}\cdot\hat{\nabla}v_jdx = &\int_\Omega\left(\hat{\mu}^2-\hat{\rho}\right)\hat{\phi}v_jdx \nonumber\\
    &- \left(\hat{\mu}^2-1\right)\int_\Omega\left(3\hat{\phi}_k^2\hat{\phi} - 2\hat{\phi}_k^3\right)v_jdx\;.
    \label{integral_eom_taylor_expanded}
\end{align}
This can be written as a linear matrix relation, 
\begin{equation}
    \mathbf{M}\mathbf{U} = \mathbf{b}\;,
\end{equation}
where $\mathbf{U}$ is a vector with elements $U_i$. 

In the finite element method, the domain of the problem $\Omega$ is divided into `cells' that are defined by their vertices $P_i$. We can decompose the field using the basis functions $e_i(\underline{x})$, such that 
\begin{equation}
    \hat{\phi} = \sum_i\hat{\Phi}_i\hat{e}_i\;,
\end{equation}
where $\hat{\Phi}_i=\hat{\Phi}(P_i)$. Let $\mathbf{M}$ be a matrix with elements
\begin{equation}
    M_{ij} = \int_\Omega \hat{\nabla}\hat{e}_i\cdot\hat{\nabla}v_jdx\;,
\end{equation}
then
rewriting the terms in the equation of motion, equation (\ref{integral_eom_taylor_expanded}), in terms of the basis functions $\hat{e}_i$, we have 
\begin{align}
    \alpha\mathbf{M}\mathbf{\hat{\Phi}} =& \left[\int_\Omega\hat{\mu}^2\cdot\hat{e}_i\cdot v_jdx\right]\hat{\Phi}_i - \left[\int_\Omega\hat{\rho}\cdot\hat{e}_i\cdot v_jdx\right]\hat{\Phi}_i \nonumber\\
    &- \left[\int_\Omega 3\hat{\phi}_k^2\left(\hat{\mu}^2-1\right)\cdot\hat{e}_i\cdot v_jdx\right]\hat{\Phi}_i\nonumber\\ &+ 2\left(\hat{\mu}^2-1\right)\int_\Omega \hat{\phi_k^3}v_jdx\;.
\end{align}
Now let $\mathbf{P}$ be a vector with elements 
\begin{equation}
    P_{ij} = \int_\Omega\hat{\rho}\cdot\hat{e}_i\cdot v_jdx\;,
\end{equation}
and let $\mathbf{B}_k$ and $\mathbf{C}_k$ be matrices with elements
\begin{subequations}
\begin{equation}
    \left[B_k\right]_{ij} = \int_\Omega \hat{\phi}_k^2\cdot\hat{e}_i\cdot v_jdx\;,
\end{equation}
\begin{equation}
    \left[C_k\right]_{ij} = \int_\Omega \hat{\phi}_k^3\cdot v_jdx\;.
\end{equation}
\end{subequations}
In terms of these matrices, the symmetron equation of motion is
\begin{align}
    \left[\alpha\mathbf{M} + \mathbf{P} + 3\left(\hat{\mu}^2-1\right)\mathbf{B}_k - \hat{\mu}^2\mathbb{1}\right]\hat{\mathbf{\Phi}} = 2\left(\hat{\mu}^2-1\right)\mathbf{C}_k\;,
    \label{sym_eom_matrix}
\end{align}
where $\mathbb{1}$ is the identity matrix. 

\section{Evolving the motion of the test particle}
\label{app:boot}
We will be interested in solving for the motion of matter particles under the influence of a symmetron fifth force.  We solve for their motion using a leapfrog algorithm. This algorithm consists of an initial desynchronization of the velocity:
\begin{equation}
        v_{1/2} = v - \frac{1}{2}a\text{d}t\;,
    \end{equation}
where $a$ is the acceleration due to the symmetron field. In terms of the gradient of the scalar field $\hat{\nabla}\hat{\phi}$, the acceleration $a$ is given by
\begin{equation}
    a = -\frac{c^2}{L}\frac{\mu^2}{\lambda M^2}\hat{\phi}\hat{\nabla}\hat{\phi}\;.
\end{equation}
Inside a loop over time, the position and velocity of the test particle are updated via
\begin{subequations}
    \begin{equation}
        v_{1/2} = v_{1/2} + a\text{d}t\;,
    \end{equation}
    \begin{equation}
        x = x + v_{1/2}\text{d}t\;,
    \end{equation}
    \begin{equation}
        v = v_{1/2} - \frac{1}{2}a\text{d}t\;,
    \end{equation}
    \end{subequations}
after which the velocity is resynchronized with
\begin{equation}
        v = v_{1/2} - \frac{1}{2}a\text{d}t\;.
    \end{equation}

\bibliography{arxiv_draft}
\end{document}